\documentclass[prb,aps, superscriptaddress,twocolumn]{revtex4-1}
\usepackage{amsmath,amssymb}
\usepackage{graphicx}
\usepackage{wasysym}
\usepackage{amsfonts}
\usepackage{bm}
\usepackage{enumerate}
\usepackage{color}
\usepackage[resetlabels]{multibib}
\usepackage{epstopdf}
\usepackage{latexsym}
\usepackage[breaklinks,colorlinks = true,linkcolor = red,urlcolor=cyan,citecolor=red]{hyperref}
\usepackage[caption=false,singlelinecheck=false]{subfig}

\usepackage{times}
\newcommand{\bea}{\begin{eqnarray}}
\newcommand{\eea}{\end{eqnarray}}
\newcommand{\be}{\begin{eqnarray}}
\newcommand{\ee}{\end{eqnarray}}
\newcommand{\bw}{\begin{widetext}}
\newcommand{\ew}{\end{widetext}}

\newcommand\numberthis{\addtocounter{equation}{1}\tag{\theequation}}

\begin{document}
\title{Subharmonic Fidelity Revival in a Driven PXP model}
\author{HaRu K. Park}
\email{haru.k.park@kaist.ac.kr}
\affiliation{Department of Physics, Korea Advanced Institute of Science and Technology, Daejeon, 34141, Korea}
\author{SungBin Lee}
\email{sungbin@kaist.ac.kr}
\affiliation{Department of Physics, Korea Advanced Institute of Science and Technology, Daejeon, 34141, Korea}
\date{\today}
\begin{abstract}
The PXP model hosts a special set of nonergodic states, referred to as quantum many-body scars. One of the consequences of quantum scarring is the periodic revival of the wave function fidelity.
It has been reported that quantum fidelity revival occurs in the PXP model for certain product states, and periodic driving of chemical potential can enhance the magnitude of quantum revival, and can even change the frequencies of revival showing the subharmonic response. Although the effect of the periodic driving in the PXP model has been studied in the limit of certain perturbative regimes, the general mechanism of such enhanced revival and frequency change has been barely studied.
  In this work, we investigate how periodic driving in the PXP model can systematically control  the fidelity revival. Particularly, focusing on the product state so called  a  N{\'e}el state,  we analyze the condition of driving to enhance the magnitude of revival or change the frequencies of revival.
To clarify the reason of such control, we consider the similarities between the PXP model and the free spin-$1/2$ model in graph theoretical analysis, 
and show that the quantum fidelity feature in the PXP model is well explained by the free spin-$1/2$ model.
In addition, under certain limit of  the driving parameters, analytic approach to explain the main features of the fidelity revival is also performed.  
Our results give an insight of the scarring nature of the periodically driven PXP model and pave the way to understand their (sub-)harmonic responses and controls. 
 \end{abstract}

%\pacs{}
%\keywords{}

\maketitle

\section{Introduction}
Eigenstate Thermalization Hypothesis(ETH)\cite{PhysRevE.50.888, PhysRevA.43.2046, rigol2008thermalization,doi:10.1126/science.aaf6725} is a key concept explaining the thermalization of the quantum many-body system. Recently, beyond the quantum thermalization, the system which strongly violates the ETH has been actively studied, such as integrable systems and many-body localization. Here, the "strong violation" of the ETH means every eigenstate breaks the ergodicity and never get thermalized. There are also the systems which "weakly violate" the ETH, implying only a small portion of the eigenstates violates the ergodicity while all the other states get thermalized. In particular, quantum many-body scarring(QMBS) systems are the examples which weakly violates ETH, containing a small number of highly excited non-thermal eigenstates, called scar eigenstates\cite{PhysRevLett.53.1515,serbyn2021quantum}. 
Such scar eigenstates show exotic physical behavior compared to the thermal Gibbs state. For example, while the thermal Gibbs state predicts the entanglement entropy proportional to the volume of subsystem in the middle of the spectrum, the entanglement entropy of the scar eigenstate scales proportionally to the area of the subsystem.
Because the QMBS often appear in a tower of scar states where a set of states with equidistant energy spacing exists\cite{moudgalya2021quantum}, if the initial product state is a superposition of the scar states, then the system exhibits the perfect revival. Conversely, it has been also shown that the persistent revival of product state implies QMBS\cite{PhysRevB.101.205107}. Hence, it is important to observe the fidelity revival as evidence for the QMBS in experiments.

In recent experiments of a Rydberg atom simulator, it has been observed that certain product states show persistent, though imperfect revival under van der Waals interaction. This indicates the presence of a QMBS in the system\cite{bernien2017probing,doi:10.1126/science.abg2530,turner2018weak}. Taking the extreme limit of strong van der Waals interaction in the Rydberg blockade of the Rydberg atom chain gives rise to the so-called PXP model, which has a Hilbert space projected onto the states with no neighboring excited states. The QMBS structure of the PXP model has been actively studied theoretically, including the report that it also shows the imperfect revival\cite{PhysRevB.98.155134}. There exist many attempts to reach high fidelity revival, for instance, by enhancing its weakly broken $SU(2)$ symmetry\cite{PhysRevLett.122.220603,PhysRevB.101.165139}.
There is another way which enhances the quantum revival of the states in the PXP model: periodic driving. In most of the cases, periodic driving induces thermalization of the scar states, and thus destructs the fidelity revival. However, recent experimental and theoretical studies show\cite{doi:10.1126/science.abg2530, PhysRevB.106.104302} that periodic driving with certain amplitudes surprisingly enhances the fidelity revival. 
Furthermore, it has also been observed that the subharmonic response of fidelity revival exists with doubled period compared to the driving mode. This kind of subharmonic response breaks the discrete time translational symmetry of the driving mode, and hence get attention as a time-version of crystalline order, called a discrete time-crystal (DTC) which is also a recently studied subject\cite{PhysRevLett.117.090402,PhysRevLett.118.030401}.  Although earlier research had demonstrated this subharmonic fidelity revival in the limit of  large driving amplitude and high frequency\cite{PhysRevLett.127.090602}, the general mechanism of such subharmonic revival has not been explored well.
 
In this paper, we study the periodically driven PXP model, focusing on the subharmonic fidelity revivals. In Section \ref{sec:2}, we introduce the PXP model with square pulse driving modes. By calculating the average fidelity and the Fourier components of the fidelity signal, we study the conditions under which the fidelity is enhanced and subharmonic response occurs. Then, based on the similarity of the Hamiltonian adjacent graph between PXP model and free spin model\cite{PhysRevB.105.245137}, we show that they can be explained by the free spin-$1/2$ model with the same driving which is exactly solvable.
In Section \ref{sec:3}, we introduce an analytic approach for the driven PXP model. 
Within perturbative analysis, we derive the driving conditions for subharmonic responses in the driven PXP model and discuss their applications.
In Section \ref{sec:4}, we summarize our works and suggest interesting future directions.

\section{PXP Model with square pulse drive}\label{sec:2}

In this section, we first introduce the static PXP model and  then represent how the fidelity revival is controlled under periodic driving. Particularly, based on the graph theoretical similarity between PXP model and free spin-$1/2$ model, we argue that various phenomena in the driven PXP model, such as the revival enhancement and (sub-) harmonic response, can be explained by the free spin-$1/2$ model.

The PXP model, describing Rydberg atoms with strong interaction, is represented as following,
\begin{equation}\label{eq:pxp}
H_{\textrm{PXP}} = \Omega \sum_j P_{j-1} X_j P_{j+1}.
\end{equation}Here, the spin at each site consists of two states, $|0\rangle$ and $|1\rangle$, which represent a ground state and an excited state,  respectively. 
$X_j=|0_j\rangle\langle 1_j|+|1_j\rangle \langle 0_j|$ is the Pauli $x$ matrix at site $j$, and $P_j=|0_j\rangle \langle 0_j|$ is the projection operator for ground state at site $j$.
This Hamiltonian describes the system which prohibits the spin-flip, unless the neighboring sites are in the ground state, i.e. only the transition, $|\cdots 010\cdots \rangle \cdots \leftrightarrow |\cdots 000\cdots\rangle$, is allowed. Since this transition does not generate or annihilate excited states in any two neighboring sites, one can exclude the states where the consecutive neighbors are  excited.

Although the PXP model is a non-integrable chaotic system\cite{turner2018weak}, there are certain product states which show non-ergodicity and decent revival under time evolution, such as $|\mathbb{Z}_2\rangle \equiv |0101\cdots 01\rangle$ called a  N{\'e}el state. 
The non-ergodic property of such product states is unique, in a sense that the number of them increases linearly with the system size, whereas, other ergodic product states show exponential increase with the system size. It has been understood that their fidelity revival is originated from the quantum scarring, i.e., the product state with short-time revival is a linear combination of scar eigenstates with equivalent energy spacing\cite{moudgalya2021quantum}. 
The $|\mathbb{Z}_2\rangle$ state is mainly composed of the quantum scar states with almost equal energy spacing. However, because their energy spacing is not perfectly even, it has been pointed out that the fidelity revival of the $|\mathbb{Z}_2\rangle$ state is also imperfect\cite{PhysRevB.98.155134}.

There have been many suggestions to adjust the system enhancing this imperfect revivals. As one promising way, it has been studied in both theoretically and experimentally to the addition of cosine modulation. This modulation plays a role of controlling chemical potential to the PXP model, and can enhance the fidelity revival of the $|\mathbb{Z}_2\rangle$ state or even induce subharmonic responses in certain driving condition\cite{doi:10.1126/science.abg2530,PhysRevB.106.104302}. 
However, the systematic ways to find such driving have not been studied in detail, which is  the focus of this study.

The periodically driven PXP model is represented as following.
\begin{equation}\label{eq:pxp_driven}
H(t)=H_{\textrm{PXP}} + \Delta_{\textrm{sq}}(t) \sum_{j} n_j,
\end{equation}
where $H_{\textrm{PXP}}$ is defined in Equation \ref{eq:pxp} and the second term represents the periodic driving, where $n_j=|1_j\rangle \langle 1_j|$ counts the number of excited states on each site. In terms of the periodic driving, we adopt the square pulse driving protocol as also introduced in earlier studies for analysis. It shares a very similar fidelity profile with the cosine driving case and thus well explains the experiments\cite{PhysRevB.106.104302}. 
The square pulse driving protocol $\Delta_{\textrm{sq}}(t)$ within the period $T$ is defined as,
\begin{equation}\label{eq:driving_protocol}
\Delta_{\textrm{sq}}(t)=\begin{cases}
\Delta_0 + \Delta_m,& 0\leq t \leq T/4\\
\Delta_0 - \Delta_m,& T/4<t\leq 3T/4\\
\Delta_0 + \Delta_m,& 3T/4<t<T.
\end{cases}
\end{equation}
This corresponds to the periodic driving with frequency $\omega_0=2\pi/T$, average chemical potential $\Delta_0$, and driving amplitude $\Delta_m$.

\begin{figure*}[ht]
  \includegraphics[width=0.9\textwidth]{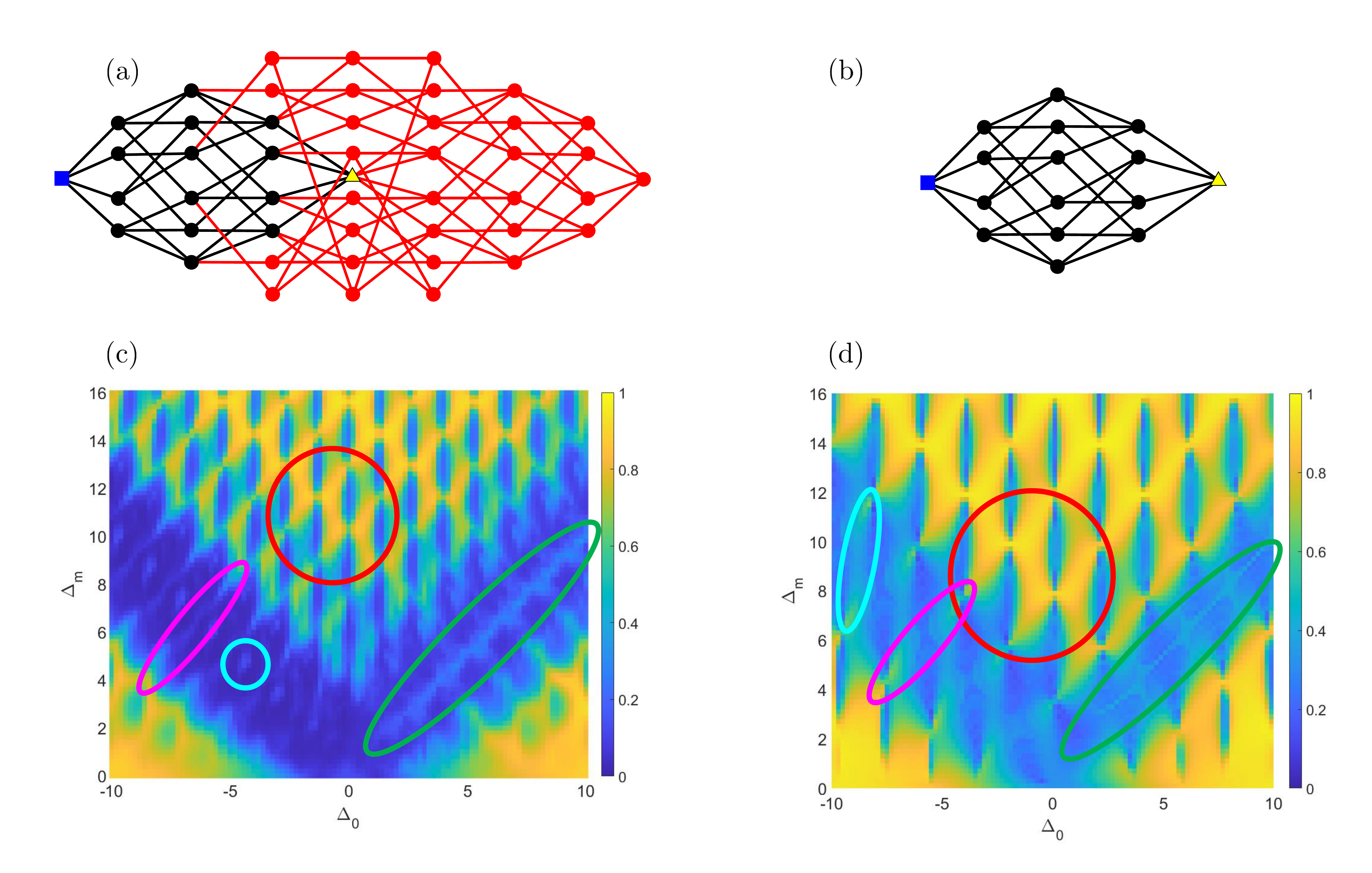}
  \caption{(Top) The Hamiltonian adjacency graph of (a) PXP model for $L=8$ and (b) free spin $1/2$ model for $L=4$. The blue square and yellow triangle in Figure (a) represent the states $|01010101\rangle$ and $|00000000\rangle$ respectively, and the blue square and yellow triangle in Figure (b) represent the states $|1111\rangle$ and $|0000\rangle$ respectively. (Bottom) Average fidelity, $\overline{F}(0)$, with initial state $|\mathbb{Z}_2\rangle$ through $10T$ time domain. Figure (c) shows $\overline{F}(0)$ of PXP model with the driving period $T=4.788$ for system size $L=12$; Figure (d) shows $\overline{F}(0)$, of free spin-$1/2$ model with driving period $T=\pi$ for  system size $L=6$. The "butterfly" peaks are encircled by red line, "bridge" peaks are encircled by magenta line, "local" (in Figure (c)) or "steep bridge" (in Figure (d)) peaks are encircled by cyan line, and "separator" peaks are encircled by blue line.}
  \label{fig:pxp-fidelity}
\end{figure*}

Now, we introduce the wave function fidelity 
$F(t)\equiv |\langle \psi(t)|\psi(0)\rangle|^2$
to measure how the revival of the initial product state, $|\psi(0)\rangle = |\mathbb{Z}_2\rangle$, changes as the driving parameters $(\Delta_0, \Delta_m)$ are tuned. As a tool to measure the subharmonic response, we also introduce the Fourier component of the wave function fidelity defined as,
\begin{equation}\label{fidelity_fourier}
\overline{F}(\omega)=\left|\frac{1}{nT}\int_0^{nT} F(t) e^{-i \omega t}dt\right|.
\end{equation}
Later, we will discuss the three main quantities, $\overline{F}(0), \overline{F}(\omega_0)$ and $\overline{F} (\omega_0/2)$, as functions of $(\Delta_0 , \Delta_m)$: $\overline{ F}(0)$ introduces how the fidelity revival gets enhanced, and $ \overline {F}(\omega_0)$ and $\overline{F}(\omega_0/2)$ are responsible for harmonic and subharmonic responses, respectively. Note that if $|\psi(t)\rangle$ is an eigenstate of $H(t)$ with driving parameters $(\Delta_0, \Delta_m)$, then $\prod_j Z_j |\psi(t)\rangle$ with Pauli $z$ matrix $Z_j = |1_j\rangle\langle 1_j|-|0_j\rangle\langle 0_j|$ is also an eigenstate of $H(t)$ with the same driving parameters. Hence, without loss of generality, we only plot the region, $\Delta_m\geq 0$.

Before investigating the fidelity profile of the $|\mathbb{Z}_2\rangle$ state in the PXP model, let us introduce another model which shows very similar feature: the free spin-$1/2$ chain model with the same driving $\Delta_{\textrm{sq}}(t)$,
\begin{equation}\label{eq:free-spin}
H_{\textrm{free}}(t)=\sum_j X_j + \Delta_{\textrm{sq}}(t) \sum_{j} n_j.
\end{equation}
For the $|\mathbb{Z}_2\rangle$ state, the free spin-$1/2$ model and the PXP model share common features in terms of graph theoretical point of view.
To understand it, notice that the PXP model is nothing but the free spin-$1/2$ model with constraints. Hence, the graph of length $L$ PXP model, for instance, is a subgraph of the free spin-$1/2$ model with the same length $L$.
 Conversely, consider the length $L$ PXP model with even $L$. If we give the stronger constraint to the PXP model and only allow the states with $|0\rangle$ states at odd sites, then each states $|0x0y\cdots 0z\rangle$ can be mapped to the state $|xy\cdots z\rangle$ in length $L/2$ free spin-$1/2$ model, where $x,y,\cdots,z$ are either $0$ or $1$. This shows that the graph of length $L/2$ free spin-$1/2$ model is a subgraph of length $L$ PXP model.

Figure \ref{fig:pxp-fidelity}(a) shows the graph of the PXP Hamiltonian with $L=8$, and Figure \ref{fig:pxp-fidelity}(b) shows the graph of the free spin-$1/2$ Hamiltonian for $L=4$. The blue square in Figure \ref{fig:pxp-fidelity}(a) marks the N{\'e}el state $|01010101\rangle$ in PXP Hamiltonian, which corresponds to the $|1111\rangle$ state in free spin-$1/2$ model also marked by the blue square in Figure \ref{fig:pxp-fidelity}(b). The yellow triangles in Figure \ref{fig:pxp-fidelity}(a) and (b) represent the polarized state $|00000000\rangle$ state and $|0000\rangle$ state, respectively. The vertices and edges colored in red show the difference between the two graphs and show that the graph in Figure \ref{fig:pxp-fidelity}(b) is indeed a subgraph of Figure \ref{fig:pxp-fidelity}(a).
 Despite their difference, they share a common feature particularly for the $|\mathbb{Z}_2\rangle$ state, marked by the blue square. This common feature is generally applicable  for the graphs of the length $L$ PXP model and length $L/2$ free spin-$1/2$ model. To argue it in detail, let's consider the expansion, $
\langle \mathbb{Z}_2 (t)|\mathbb{Z}_2\rangle = \langle \mathbb{Z}_2|e^{iHt}|\mathbb{Z}_2\rangle = \sum_n \frac{(it)^n}{n!}\langle H^n\rangle$
with $\langle H^n\rangle=\langle \mathbb{Z}_2|H^n|\mathbb{Z}_2\rangle$.
 In the graph theoretical point of view, $\langle H^n\rangle$ counts the number of walks with length $n$ which starts and ends at $|\mathbb{Z}_2\rangle$ vertex. Because the difference between PXP graph and free spin graph mostly occurs for the states with high Hamming distance, their difference only affects on the long walks, i.e. $\langle H^n \rangle$ with large $n$.
 Hence, we may expect the similar behavior in $F(t)$ for a short time scale $t$ in between PXP model and free spin model. Later, we will show that it is indeed the case by comparing calculation of the fidelity revival on both PXP model and free spin model.
 For completeness we note that this argue is not applicable for ergodic initial states: for example the polarized state $|0_p\rangle \equiv |0000\cdots00\rangle$, marked by the yellow triangle, shows a large difference in graphs even for the nearest neighbors, indicating the different fidelity profile between PXP model and free spin model. 

In the presence of driving, one may also suggest the similarities between PXP model and free spin-$1/2$ model, with respect to the graph theoretical approach.
Figure \ref{fig:pxp-fidelity}(c) and \ref{fig:pxp-fidelity}(d) show the values of $\overline{F}(0)$ for PXP model and free spin-$1/2$ model respectively, as functions of driving parameters $\Delta_0$ and $\Delta_m$. 
As discussed earlier, $\overline{F}(0)$ indicates the enhancement of the fidelity revival.
Indeed, Figures \ref{fig:pxp-fidelity}(c) and \ref{fig:pxp-fidelity}(d) show very similar features up to scale. 
For calculation,  we choose the periodicity $T_0=4.788$ for the PXP model and $T_f=\pi$ for the free spin model respectively, which are optimized values for the fidelity revival observed in the static cases. The system size $L=12$ is chosen with the time range $[0,10T_0]$ for the PXP model, and $L=6$ with the time range $[0, 10T_f]$ for the free spin model. 

In Figure \ref{fig:pxp-fidelity}(c) and \ref{fig:pxp-fidelity}(d), we point out several common features as following. 
We first note the "butterfly"-shaped peaks on top of each figure, marked by a red circle, and high average fidelity region on the lower left and right side. 
Next, there is a wide $V$-shaped region having relatively small values of $\overline{F}(0)$ in-between, with the following substructures: On the left side in both Figures \ref{fig:pxp-fidelity}(c) and (d), the butterfly peaks are connected to the lower left region by some "bridges", marked by magenta lines. In addition, there are "local" peaks between the bridges marked by cyan line in Figure \ref{fig:pxp-fidelity}(c), but instead there are "steep bridge" peaks in Figure \ref{fig:pxp-fidelity}(d).
Later, we will explain that they indicate the same phenomena. On the right side in both Figures \ref{fig:pxp-fidelity}(c) and (d), there are long and thin "separator" peak marked by green line, which separates the butterfly peaks and the lower right region.

\begin{figure}[t]
  \subfloat[]{\label{fig:pxp-onefreq}\includegraphics[width=0.23\textwidth]{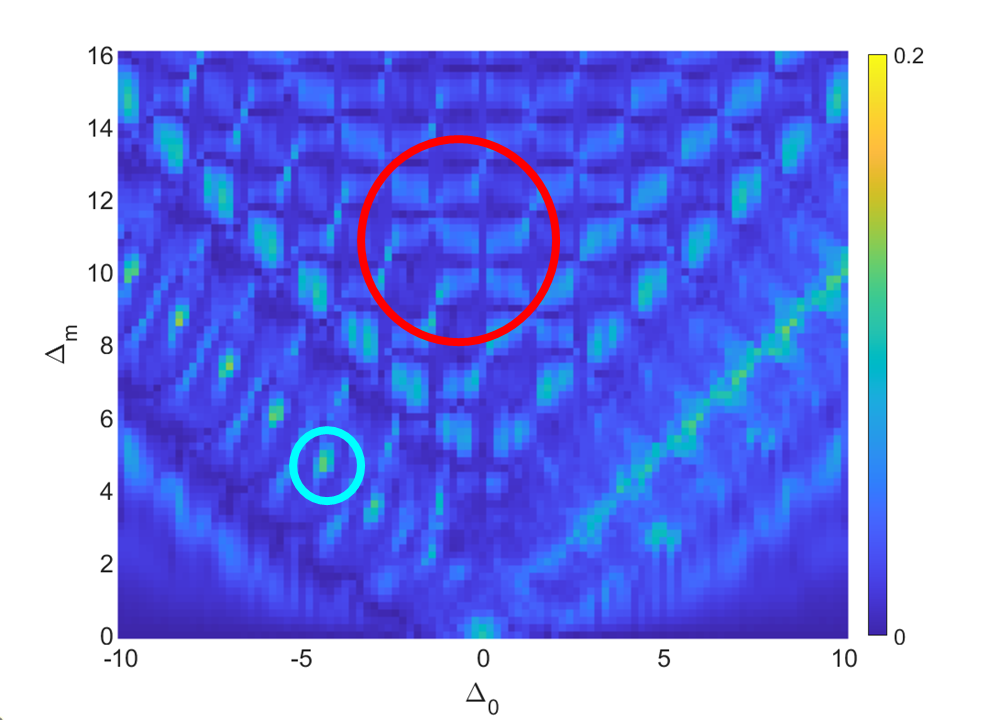}}
  \subfloat[]{\label{fig:pxp-halffreq}\includegraphics[width=0.23\textwidth]{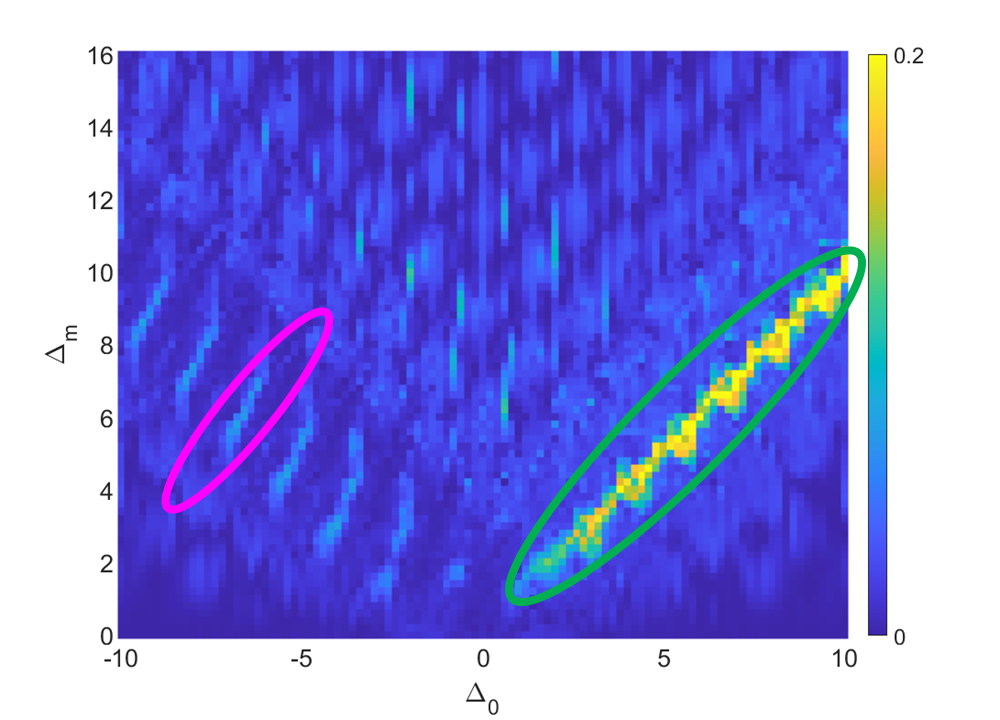}}
  \caption{Frequency profile of the fidelity $F(t)$, \ref{sub@fig:pxp-onefreq} $\overline{F}(\omega_0)$ and \ref{sub@fig:pxp-halffreq} $\overline{F}(\omega_0/2)$ with initial state $|\mathbb{Z}_2\rangle$, for the PXP model with driving period $T=4.788$ and system size $L=12$. The "butterfly" peaks are encircled by red line, "bridges" peaks are encircled by magenta line, "local" peaks are encircled by cyan line, and "separator" peaks are encircled by blue line. }
  \label{fig:pxp-oneandhalffreq}
\end{figure}

To determine the origin of these peaks, the frequency profiles of the fidelity are investigated.   
Figures \ref{fig:pxp-onefreq} and \ref{fig:pxp-halffreq} plot the Fourier component values of the fidelity for the $|\mathbb{Z}_2\rangle$ state in the PXP model, $\overline{F}(\omega_0)$ and $\overline{F}(\omega_0/2)$, with $\omega_0=2\pi/T_0$.
By comparing them with Figure \ref{fig:pxp-fidelity}(c), one can conclude that the butterfly peaks (marked by red line) and the local peaks (cyan line) represent harmonic revivals, whereas, the bridges (magenta line) and separators (green line) represent the subharmonic revivals.
  Notice that the bridges and the separators also appear in Figure \ref{fig:pxp-onefreq}. However, this does not imply that they are harmonic responses, since the subharmonic response with nonzero $\overline{F}(\omega_0/2)$ also has finite values of $\overline{F}(\omega_0)$. Therefore, Figure \ref{fig:pxp-halffreq} is a direct evidence, showing the subharmonic response indeed occurs due to the driving.

\begin{figure}[t]
  \subfloat[]{\label{fig:free-spin-onefreq}\includegraphics[width=0.23\textwidth]{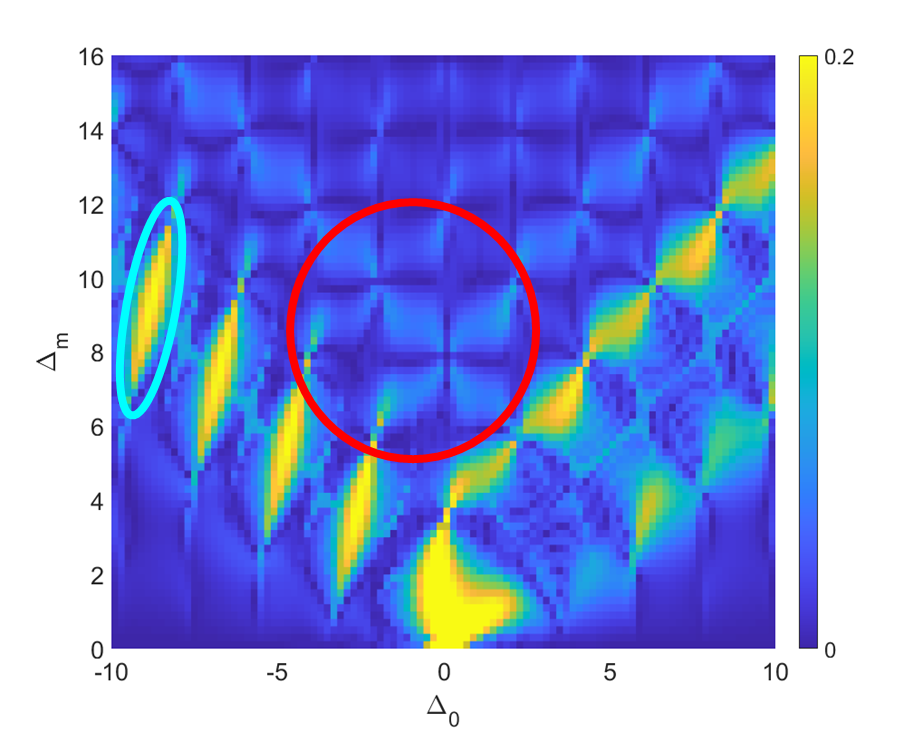}}
  \subfloat[]{\label{fig:free-spin-halffreq}\includegraphics[width=0.23\textwidth]{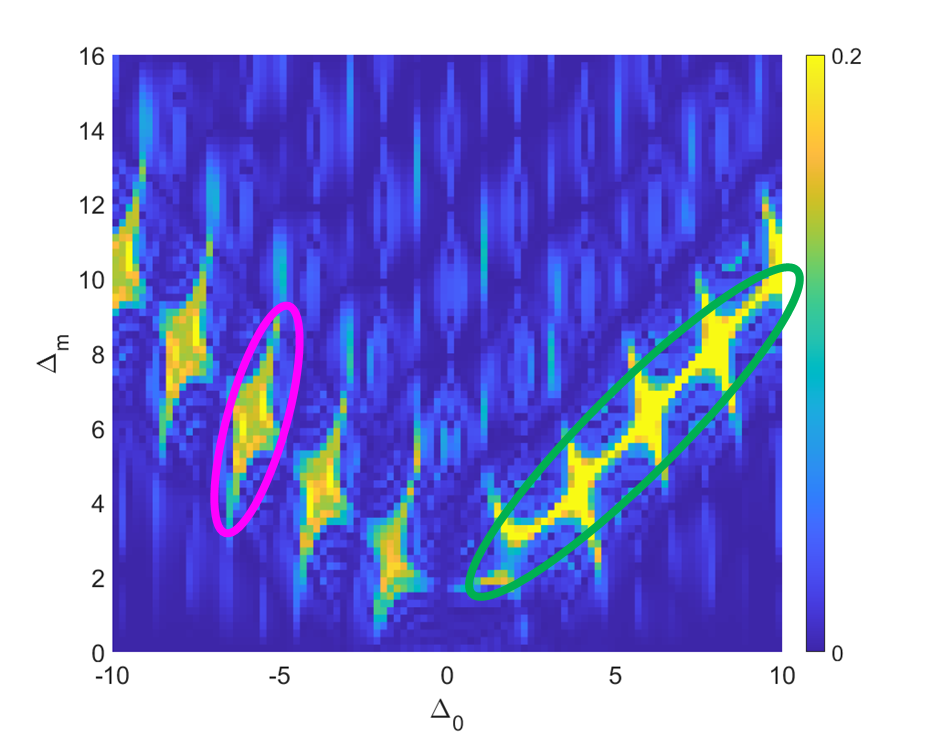}}
  \caption{Frequency profile of the fidelity $F(t)$, \ref{sub@fig:free-spin-onefreq} $\overline{F}(\omega_0)$ and \ref{sub@fig:free-spin-halffreq} $\overline{F}(\omega_0/2)$ with initial state $|\mathbb{Z}_2\rangle$, for the free spin model with driving period $T=\pi$ and system size $L=12$. The "butterfly" peaks are encircled by red line, "bridges" peaks are encircled by magenta line, "steep bridge" peaks are encircled by cyan line, and "separator" peaks are encircled by blue line.}
  \label{fig:free-spin-oneandhalffreq}
\end{figure}

For comparison, we also investigate the free spin-$1/2$ model case. Figures \ref{fig:free-spin-onefreq} and \ref{fig:free-spin-halffreq} plot the values of $\overline{F}(\omega_f)$ and $\overline{F}(\omega_f/2)$ for the free-spin model, showing harmonic and subharmonic responses respectively, with $\omega_f=2\pi/T_f=2$. The butterfly peaks (marked by red line) and the steep bridge peaks (cyan line) in Figure \ref{fig:free-spin-onefreq} again shows the harmonic response, while the bridge peaks (magenta line) and the separator peaks on the right side (green line) in Figure \ref{fig:free-spin-halffreq} shows the subharmonic response. Indeed, these features are consistent with the case of the PXP model which is explained earlier (see Figure \ref{fig:pxp-oneandhalffreq}).

\section{Perturbative and exact calculations on the models}\label{sec:3}
Until now, we have shown that the periodic driving of the PXP model can induce the subharmonic responses of the $|\mathbb{Z}_2\rangle$ state fidelity and have interpreted them based on the graph theoretical similarities with the free spin-$1/2$ model. In the following, for more rigorous argument, alternative analytic approaches are presented to understand subharmonic responses and to determine the optimal driving conditions. Since our focus lies on the subharmonic response of driven PXP model, we perform the appropriate perturbation limit which represents the $V$-shaped region in Figure \ref{fig:pxp-fidelity}(c), where every subharmonic peaks lies on.
 We note that perturbation approach in another limit has been already performed in earlier work\cite{PhysRevLett.127.090602}. Its perturbative limit  explains the "butterfly" peaks, marked by the red circle in Figure \ref{fig:pxp-fidelity}(c). In contrast, our perturbative limit explains every peaks on the $V$-shaped region, including "bridge", "local", and "separator" peaks.

Before moving on, we redefine some notations for simplicity. Take $\Delta_\pm \equiv (\Delta_0\pm \Delta_m)/2$, $\overline{X}_j\equiv P_{j-1}X_j P_{j+1}$, and $H^{\pm}\equiv \sum_j H^\pm_j$ with $H^\pm_j \equiv \Omega \overline{X}_j +\Delta_\pm Z_j$. Notice that $H^\pm = H_{\textrm{PXP}} + 2\Delta_{\pm} \sum_j n_j - \Delta_\pm \sum_j I_j$, hence the evolution operator in the presence of square pulse driving,
\begin{equation}\label{eq:evol_mat}
U=e^{iH^+ T_0/4}e^{iH^- T_0/2}e^{iH^+ T_0/4},
\end{equation} 
is equivalent to the evolution operator of $H(t)$ in one period up to phase.

Consider the limit $\Delta_+\gg \Omega\gg \Delta_-$, which is the right side of the $V$-shaped region with low values of $\overline{F}(0)$ in Figure \ref{fig:pxp-fidelity}. We will show that this limit always gives subharmonic response. In this limit, we can approximate $H^+\simeq \Delta_+ \sum_j Z_j$ and $H^-\simeq \Omega H_{\textrm{PXP}}$ taking the leading terms. Because our product state is an eigenstate of $H^+$, if we calculate $\langle \mathbb{Z}_2'|U|\mathbb{Z}_2\rangle$, where $\mathbb{Z}_2'\equiv |1010\cdots 10\rangle$ is a translated N{\'e}el state, one can easily show $\langle \mathbb{Z}_2'|U|\mathbb{Z}_2\rangle \simeq \langle \mathbb{Z}_2'|e^{iH_{\textrm{PXP}}T_0/2}|\mathbb{Z}_2\rangle $ up to phase. 
For $L=12$ case, this value is quite large $\simeq 0.9658$.
This results in the $2T$-periodic revival with high lower bound of $|\langle \mathbb{Z}_2|U^2|\mathbb{Z}_2\rangle|$, satisfying, 
\begin{equation}\label{eq:subharmonic_revival_bound}
|\langle \mathbb{Z}_2|U^2|\mathbb{Z}_2\rangle|\geq |\langle \mathbb{Z}_2|e^{iH_{\textrm{PXP}}T_0/2}|\mathbb{Z}_2'\rangle|^2,\quad \Delta_+\gg \Omega\gg \Delta_-.
\end{equation} 
See Appendix \ref{appendix_a} for detailed proof. Thus, one can claim the persistent subharmonic revival indeed occurs in $\Delta_+\gg \Omega\gg \Delta_-$.
We also show that this revival is robust even with $\mathcal{O}(\Omega)$ order terms in Appendix \ref{appendix_a}.

Now consider the limit $\Delta_- \gg \Omega \gg \Delta_+$, which is the left side of the $V$-shaped region with low values of $\overline{F}(0)$ in Figure \ref{fig:pxp-fidelity}. Similar analysis with the case $\Delta_+\gg \Omega\gg \Delta_-$ leads $H^+\simeq \Omega H_{\textrm{PXP}}$ and $H^-\simeq \Delta_- \sum_j Z_j$. 
We focus on the driving conditions for the parameters  $\Delta_-=\frac{n\pi}{T_0}=\frac{n\omega_0}{2}$, and our aim is to show that for even $n$ case the fidelity presents subharmonic response and for odd $n$ case the fidelity presents harmonic response. First, let $n=2k$ be even. In this case, $e^{iH^- T/2} = 1$, and thus one again achieve $\langle \mathbb{Z}_2'|U|\mathbb{Z}_2\rangle\simeq \langle \mathbb{Z}_2'|e^{H_{\textrm{PXP}}T_0/2}|\mathbb{Z}_2\rangle$. Thus, we get the very similar result with Equation \ref{eq:subharmonic_revival_bound},
\begin{align*}\label{eq:analytic_result_2}
|\langle \mathbb{Z}_2|U^2|\mathbb{Z}_2\rangle|\geq |\langle \mathbb{Z}_2|&e^{iH_{\textrm{PXP}}T_0/2}|\mathbb{Z}_2'\rangle|^2,\\
&\quad \Delta_-=k\omega_0 \gg \Omega\gg \Delta_+,\numberthis
\end{align*} 
showing the subharmonic response mainly occurs at $\Delta_- = k\omega_0$ for integer $k$'s. 
On the other hand, for odd $n=2k+1$'s, $e^{iH^- T_0/2}=\prod_j Z_j$. 
Using the anti-commutation relation between $\prod_j Z_j$ and $H_{\textrm{PXP}}$, the evolution operator is represented as,
\begin{align*}
U&\simeq e^{iH_{\textrm{PXP}}T_0/4}\left(\prod_j Z_j\right) e^{iH_{\textrm{PXP}}T_0/4}\\
&= \prod_j Z_j e^{-iH_{\textrm{PXP}}T_0/4} e^{iH_{\textrm{PXP}}T_0/4} = \prod_j Z_j,\numberthis
\end{align*}
and this results in,
\begin{equation}\label{eq:analytic_result_3}
|\langle \mathbb{Z}_2|U|\mathbb{Z}_2 \rangle| \simeq 1,\quad \Delta_-=\left(k+\frac{1}{2}\right)\omega_0 \gg \Omega\gg \Delta_+.\numberthis
\end{equation}
Thus, the harmonic response mainly occurs for $\Delta_-=(k+1/2)\omega_0$ region. We again show that this revival is robust up to $\mathcal{O}(\Omega)$ order, see Appendix \ref{appendix_a}.

In summary of this section, our perturbative analysis provides reasonable explanation why the several peaks in $V$-shaped region are emerging in Figure \ref{fig:pxp-fidelity}. Specifically, Equation \ref{eq:subharmonic_revival_bound} explains the long diagonal "separator" peaks on the right side of $V$-shaped region, Equation \ref{eq:analytic_result_2} explains the "bridge" peaks on the left side, and Equation \ref{eq:analytic_result_3} explains the "local" peaks between bridge peaks. It is important to note that the derivation for Equations \ref{eq:subharmonic_revival_bound}, \ref{eq:analytic_result_2} and \ref{eq:analytic_result_3} can be generally applicable.

\section{Discussion and Conclusion}\label{sec:4}
In this work, we study the wave function fidelity revival on the periodically driven PXP model.
First, we show that the driving on PXP model induces various interesting responses, including subharmonic responses. Based on the graph theoretical similarities between PXP model and free spin-$1/2$ model, we have claimed and numerically confirmed that the driving condition which induces the subharmonic response in the PXP model can be captured by the free spin-$1/2$ model. Then, considering perturbative analysis, the generic driving conditions for subharmonic responses in the PXP model are derived. Our work will shed a light on the Rydberg atom simulator, studying subharmonic responses of the driven quantum many-body scarring systems.

As an interesting future work, 
 one may extend our studies with finite van der Waals interaction, and explore the conditions of the subharmonic revival as  interaction changes.
Since the strength of the van der Waals interaction is determined by the distance between the two Rydberg atoms $r$ as $\sim \frac{1}{r^6}$,\cite{bernien2017probing}, one could control the atom distance to tune their interaction strength and track the revival property of the $|\mathbb{Z}_2\rangle$ state. One can also consider the effect of further neighbor van der Waals interactions, and explore how the fidelity revival condition changes, which we will leave as a future work.

%%%%%%%%%%%%%%%%%%%%%%%%%%%%%%%%%%%%%%%%%%%%%%%%%%%%%%
% Acknowledgments
%%%%%%%%%%%%%%%%%%%%%%%%%%%%%%%%%%%%%%%%%%%%%%%%%%%%%%

\begin{acknowledgments}
\noindent	
{\em Acknowledgments.---}
We thank Junmo Jeon for valuable discussions. This work is supported by National Research Foundation Grant (No. 2020R1A4A3079707, No. 2021R1A2C1093060),).

\end{acknowledgments}

\appendix
\section{Bound of the (sub)harmonic revival on PXP model}\label{appendix_a}
In this section, we show the harmonic and subharmonic revival is stable under small $\Omega$ values up to first order, which is discussed in Section \ref{sec:3}. We show the Equations \ref{eq:subharmonic_revival_bound}, \ref{eq:analytic_result_2} and \ref{eq:analytic_result_3} still holds if we include the $\mathcal{O}(\Omega)$ terms.

We first consider the condition $\Delta_+\gg \Omega\gg \Delta_-$. In this case, because for different sites $j\neq k$, we have
\begin{equation}
[H^+_j, H^+_k] \sim \mathcal{O}(\Omega^2),
\end{equation}
hence we may write
\begin{equation}\label{eq:deduce_start}
e^{iH^+ T_0/4}\simeq \prod_j e^{i H_j^+ T_0/4} = \prod_j e^{i(\Omega \overline{X}_j +\frac{\Delta_+}{2}Z_j)T_0/4}
\end{equation}
up to $\mathcal{O}(\Omega)$ order. This can be expanded to cosine and sine functions in $\mathcal{O}(\Omega)$ order, achieving
\begin{equation}\label{eq:deduce_middle}
e^{iH_j^+ T_0/4} \simeq \cos \frac{\Delta_+ T_0}{4}+i\sin \frac{\Delta_+ T_0}{4} \left( \frac{\Omega}{\Delta_+} \overline{X}_j + Z_j\right).
\end{equation}
If we product all the $e^{iH_j^+ T_0/4}$ terms and left only the $\mathcal{O}(\Omega)$ order terms, then we finally get
\begin{align*}
&e^{iH^+ T_0/4}\\
& \simeq e^{i\Delta_+ \sum_j Z_j T_0/4}\left(1+i\sin \frac{\Delta_+ T_0}{4}\sum_j e^{-i\Delta_+ Z_j T_0/4} \frac{\Omega}{\Delta_+}\overline{X}_j\right)\\
&=\left(1+i\sin \frac{\Delta_+ T_0}{4}\sum_j\frac{\Omega}{\Delta_+}\overline{X}_j e^{-i\Delta_+ Z_j T_0/4}\right)e^{i\Delta_+ \sum_j Z_j T_0/4}.\numberthis
\end{align*}

To calculate the subharmonic response, we consider the value $\langle\mathbb{Z}_2'|U|\mathbb{Z}_2\rangle$: if this value is large enough then it guarantees the $2T$-periodic revival with $|\langle \mathbb{Z}_2|U^2|\mathbb{Z}_2\rangle|\geq |\langle \mathbb{Z}_2'|U|\mathbb{Z}_2\rangle|^2$, because
\begin{align*}\label{eq:lowerbound_t_to_2t}
\langle \mathbb{Z}_2|U^2|\mathbb{Z}_2\rangle&= |\langle \mathbb{Z}_2 |U|\mathbb{Z}'_2\rangle|^2 + \sum_i |\langle \mathbb{Z}_2 |U| \psi_i\rangle|^2\\
&\geq |\langle \mathbb{Z}_2 |U|\mathbb{Z}'_2\rangle|^2,\numberthis
\end{align*}
where $\{\psi_i\}$ are the basis of the Hilbert space orthogonal to $|\mathbb{Z}_2'\rangle$. Now because we are considering $\Omega\gg \Delta_-$ limit, we ignore $\Delta_-$, giving $e^{iH^- T_0/2}\simeq e^{iH_{\textrm{PXP}}T_0/2}$, then we have
\begin{align*}
&\langle \mathbb{Z}_2'|U|\mathbb{Z}_2\rangle\simeq \langle \mathbb{Z}_2'| e^{iH_{\textrm{PXP}}T_0/2} |\mathbb{Z}_2\rangle\\
&+i\frac{\Omega}{\Delta_+} \sin\frac{\Delta_+ T_0}{4}\left[\langle \mathbb{Z}_2'|e^{iH_{\textrm{PXP}}T_0/2} \sum_j \overline{X}_j e^{-i\Delta_+ Z_j T_0/4} |\mathbb{Z}_2\rangle\right]\\
&+i\frac{\Omega}{\Delta_+} \sin\frac{\Delta_+ T_0}{4}\left[\langle \mathbb{Z}_2'|\sum_j  e^{-i\Delta_+ Z_j T_0/4} \overline{X}_j e^{iH_{\textrm{PXP}}T_0/2} |\mathbb{Z}_2\rangle\right].\numberthis
\end{align*}
For the second term, observe that
\begin{equation}
\sum_j \overline{X}_j e^{-i\Delta_+ Z_j T_0/4} |\mathbb{Z}_2\rangle = e^{-i\Delta_+ Z_j T_0/4}\sum_j \overline{X}_j |\mathbb{Z}_2\rangle,
\end{equation}
because there are always excited states between two ground states. Hence we get
\begin{align*}
\langle \mathbb{Z}_2'|\sum_j  e^{-i\Delta_+ Z_j T_0/4} \overline{X}_j &e^{iH_{\textrm{PXP}}T_0/2} |\mathbb{Z}_2\rangle \\
&= \langle \mathbb{Z}_2'|e^{i H_{\textrm{PXP}}T_0/2} H_{\textrm{PXP}}|\mathbb{Z}_2\rangle\\
&= \partial_t \left.\langle \mathbb{Z}_2'|e^{i H_{\textrm{PXP}} t} |\mathbb{Z}_2\rangle\right\rvert_{t=T_0/2}
\end{align*}
We numerically check that $\langle \mathbb{Z}_2'|e^{iH_{\textrm{PXP}}}|\mathbb{Z}_2\rangle$ becomes maximized at $t=T_0/2$, and hence conclude this term vanishes. Arguing similar for the third term, we get
\begin{equation}\label{eq:appendix_result_1}
\langle \mathbb{Z}_2'|U|\mathbb{Z}_2\rangle\simeq \langle \mathbb{Z}_2'| e^{iH_{\textrm{PXP}}T_0/2} |\mathbb{Z}_2\rangle,
\end{equation}
showing the persistent subharmonic revival because $|\langle \mathbb{Z}_2'|e^{iH_{\textrm{PXP}}T_0/2}|\mathbb{Z}_2\rangle|$ can be taken high enough: for example, $|\langle \mathbb{Z}_2'|e^{iH_{\textrm{PXP}}T_0/2}|\mathbb{Z}_2\rangle|\simeq 0.9658$ for $L=12$.

Now we consider the condition $\Delta_- \gg \Omega \gg \Delta_+$ region, which is the left side of the $V$-shaped low $\overline{F}(0)$ region. In this case, by the similar way achieving \ref{eq:deduce_middle} we achieve
\begin{equation}
e^{iH_j^- T_0/2} \simeq \cos \frac{\Delta_- T_0}{2}+ i \sin \frac{\Delta_- T_0}{2} \left(\frac{\Omega}{\Delta_-} \overline{X}_j + Z_j\right).
\end{equation}
Here, we specifically focus on the area where $\Delta_-=\frac{n\pi+2\eta}{T_0}$ for integer $n$'s, with small $\eta\ll L^{-1}$.

We start with even $n=2k$, giving
\begin{equation}
e^{iH_j^- T_0/2} \simeq \pm 1 \pm i \eta \left(\frac{\Omega}{\Delta_-} \overline{X}_j + Z_j\right),
\end{equation}
and
\begin{equation}
e^{iH^- T_0/2}\simeq 1\pm i\eta \sum_j \left(\frac{\Omega}{\Delta_-} \overline{X}_j + Z_j\right).
\end{equation}
Now we calculate
\begin{align*}
&\langle \mathbb{Z}_2'|U|\mathbb{Z}_2\rangle \\
&= \langle \mathbb{Z}_2' |e^{iH_{\textrm{PXP}}T_0/2} |\mathbb{Z}_2\rangle\\
&+ \pm i\eta \langle \mathbb{Z}_2' |e^{iH_{\textrm{PXP}}T_0/4}\left[\sum_j \left(\frac{\Omega}{\Delta_-} \overline{X}_j + Z_j\right)\right]e^{iH_{\textrm{PXP}}T_0/4}|\mathbb{Z}_2\rangle.\numberthis
\end{align*}
Because the second term can be squeezed by $L(1+\Omega/\Delta_-)$ again, we get
\begin{equation}\label{eq:appendix_result_2}
|\langle \mathbb{Z}_2'|U|\mathbb{Z}_2\rangle|\geq |\langle\mathbb{Z}_2'|e^{iH_{\textrm{PXP}T_0/2}}|\mathbb{Z}_2\rangle|-\eta L \left(1+\frac{\Omega}{\Delta_-}\right),
\end{equation}
and since the first term is large enough, it shows that the subharmonic response mainly occurs near $\Delta_- = \frac{2k\pi}{T_0}=k\omega_0$.

Finally, we take odd $n=2k+1$. In this case, we get
\begin{equation}
e^{iH_j^- T_0/2} \simeq \pm \eta \pm i \left(\frac{\Omega}{\Delta_-} \overline{X}_j + Z_j\right)
\end{equation}
and thus
\begin{equation}
e^{iH^- T_0/2}\simeq \pm \prod_j Z_j\left[1+\sum_j Z_j \left(\eta\pm i \frac{\Omega}{\Delta_-}\overline{X}_j\right)\right].
\end{equation}
Now by using the fact that $H_{PXP}$ and $\prod_j Z_j$ anticommutes, we get
\begin{equation}
U\simeq \pm e^{-iH_{\textrm{PXP}}T_0/4} \left[1+\sum_j Z_j\left(\eta\pm i \frac{\Omega}{\Delta_-}\overline{X}_j\right)\right]e^{iH_{\textrm{PXP}}T_0/4}.
\end{equation}
Calculating $\langle\mathbb{Z}_2|U|\mathbb{Z}_2\rangle$, the first term gives $\pm 1$. For the second term, notice that the $\eta$-dependent term squeezes $\sum_j Z_j$ operator, which gives its value at most $L$, and hence squeezed by the value $\eta L$. For the $\frac{\Omega}{\Delta_-}$ term squeezing $\sum_j Z_j \overline{X}_j = i\sum_j \overline{Y}_j$ where $\overline{Y}_j = P_{j-1} Y_j P_{j+1}$ with $Y_j$ a Pauli $y$ matrix $Y_j=i|0_j\rangle\langle 1_j| - i|1_j \rangle\langle 0_j|$, we can numerically check that this value squeezes below $\delta\simeq 0.2$. Therefore,
\begin{equation}\label{eq:appendix_result_3}
|\langle \mathbb{Z}_2|U|\mathbb{Z}_2\rangle|\leq 1-\eta L - \frac{\Omega}{\Delta_-}\delta
\end{equation}
and because $\eta L$ and $\delta$ are small enough, this represents the persistent harmonic revival.

%%%%%%%%%%%%%%%%%%%%%%%%%%%%%%%%%%%%%%%%%%%%%%%%%%%%%%%
% Bibliography
\bibliography{biblio}

\begin{thebibliography}{19}%
\makeatletter
\providecommand \@ifxundefined [1]{%
 \@ifx{#1\undefined}
}%
\providecommand \@ifnum [1]{%
 \ifnum #1\expandafter \@firstoftwo
 \else \expandafter \@secondoftwo
 \fi
}%
\providecommand \@ifx [1]{%
 \ifx #1\expandafter \@firstoftwo
 \else \expandafter \@secondoftwo
 \fi
}%
\providecommand \natexlab [1]{#1}%
\providecommand \enquote  [1]{``#1''}%
\providecommand \bibnamefont  [1]{#1}%
\providecommand \bibfnamefont [1]{#1}%
\providecommand \citenamefont [1]{#1}%
\providecommand \href@noop [0]{\@secondoftwo}%
\providecommand \href [0]{\begingroup \@sanitize@url \@href}%
\providecommand \@href[1]{\@@startlink{#1}\@@href}%
\providecommand \@@href[1]{\endgroup#1\@@endlink}%
\providecommand \@sanitize@url [0]{\catcode `\\12\catcode `\$12\catcode
  `\&12\catcode `\#12\catcode `\^12\catcode `\_12\catcode `\%12\relax}%
\providecommand \@@startlink[1]{}%
\providecommand \@@endlink[0]{}%
\providecommand \url  [0]{\begingroup\@sanitize@url \@url }%
\providecommand \@url [1]{\endgroup\@href {#1}{\urlprefix }}%
\providecommand \urlprefix  [0]{URL }%
\providecommand \Eprint [0]{\href }%
\providecommand \doibase [0]{http://dx.doi.org/}%
\providecommand \selectlanguage [0]{\@gobble}%
\providecommand \bibinfo  [0]{\@secondoftwo}%
\providecommand \bibfield  [0]{\@secondoftwo}%
\providecommand \translation [1]{[#1]}%
\providecommand \BibitemOpen [0]{}%
\providecommand \bibitemStop [0]{}%
\providecommand \bibitemNoStop [0]{.\EOS\space}%
\providecommand \EOS [0]{\spacefactor3000\relax}%
\providecommand \BibitemShut  [1]{\csname bibitem#1\endcsname}%
\let\auto@bib@innerbib\@empty
%</preamble>
\bibitem [{\citenamefont {Srednicki}(1994)}]{PhysRevE.50.888}%
  \BibitemOpen
  \bibfield  {author} {\bibinfo {author} {\bibfnamefont {M.}~\bibnamefont
  {Srednicki}},\ }\href {\doibase 10.1103/PhysRevE.50.888} {\bibfield
  {journal} {\bibinfo  {journal} {Phys. Rev. E}\ }\textbf {\bibinfo {volume}
  {50}},\ \bibinfo {pages} {888} (\bibinfo {year} {1994})}\BibitemShut
  {NoStop}%
\bibitem [{\citenamefont {Deutsch}(1991)}]{PhysRevA.43.2046}%
  \BibitemOpen
  \bibfield  {author} {\bibinfo {author} {\bibfnamefont {J.~M.}\ \bibnamefont
  {Deutsch}},\ }\href {\doibase 10.1103/PhysRevA.43.2046} {\bibfield  {journal}
  {\bibinfo  {journal} {Phys. Rev. A}\ }\textbf {\bibinfo {volume} {43}},\
  \bibinfo {pages} {2046} (\bibinfo {year} {1991})}\BibitemShut {NoStop}%
\bibitem [{\citenamefont {Rigol}\ \emph {et~al.}(2008)\citenamefont {Rigol},
  \citenamefont {Dunjko},\ and\ \citenamefont
  {Olshanii}}]{rigol2008thermalization}%
  \BibitemOpen
  \bibfield  {author} {\bibinfo {author} {\bibfnamefont {M.}~\bibnamefont
  {Rigol}}, \bibinfo {author} {\bibfnamefont {V.}~\bibnamefont {Dunjko}}, \
  and\ \bibinfo {author} {\bibfnamefont {M.}~\bibnamefont {Olshanii}},\
  }\href@noop {} {\bibfield  {journal} {\bibinfo  {journal} {Nature}\ }\textbf
  {\bibinfo {volume} {452}},\ \bibinfo {pages} {854} (\bibinfo {year}
  {2008})}\BibitemShut {NoStop}%
\bibitem [{\citenamefont {Kaufman}\ \emph {et~al.}(2016)\citenamefont
  {Kaufman}, \citenamefont {Tai}, \citenamefont {Lukin}, \citenamefont
  {Rispoli}, \citenamefont {Schittko}, \citenamefont {Preiss},\ and\
  \citenamefont {Greiner}}]{doi:10.1126/science.aaf6725}%
  \BibitemOpen
  \bibfield  {author} {\bibinfo {author} {\bibfnamefont {A.~M.}\ \bibnamefont
  {Kaufman}}, \bibinfo {author} {\bibfnamefont {M.~E.}\ \bibnamefont {Tai}},
  \bibinfo {author} {\bibfnamefont {A.}~\bibnamefont {Lukin}}, \bibinfo
  {author} {\bibfnamefont {M.}~\bibnamefont {Rispoli}}, \bibinfo {author}
  {\bibfnamefont {R.}~\bibnamefont {Schittko}}, \bibinfo {author}
  {\bibfnamefont {P.~M.}\ \bibnamefont {Preiss}}, \ and\ \bibinfo {author}
  {\bibfnamefont {M.}~\bibnamefont {Greiner}},\ }\href {\doibase
  10.1126/science.aaf6725} {\bibfield  {journal} {\bibinfo  {journal}
  {Science}\ }\textbf {\bibinfo {volume} {353}},\ \bibinfo {pages} {794}
  (\bibinfo {year} {2016})},\ \Eprint
  {http://arxiv.org/abs/https://www.science.org/doi/pdf/10.1126/science.aaf6725}
  {https://www.science.org/doi/pdf/10.1126/science.aaf6725} \BibitemShut
  {NoStop}%
\bibitem [{\citenamefont {Heller}(1984)}]{PhysRevLett.53.1515}%
  \BibitemOpen
  \bibfield  {author} {\bibinfo {author} {\bibfnamefont {E.~J.}\ \bibnamefont
  {Heller}},\ }\href {\doibase 10.1103/PhysRevLett.53.1515} {\bibfield
  {journal} {\bibinfo  {journal} {Phys. Rev. Lett.}\ }\textbf {\bibinfo
  {volume} {53}},\ \bibinfo {pages} {1515} (\bibinfo {year}
  {1984})}\BibitemShut {NoStop}%
\bibitem [{\citenamefont {Serbyn}\ \emph {et~al.}(2021)\citenamefont {Serbyn},
  \citenamefont {Abanin},\ and\ \citenamefont {Papi{\'c}}}]{serbyn2021quantum}%
  \BibitemOpen
  \bibfield  {author} {\bibinfo {author} {\bibfnamefont {M.}~\bibnamefont
  {Serbyn}}, \bibinfo {author} {\bibfnamefont {D.~A.}\ \bibnamefont {Abanin}},
  \ and\ \bibinfo {author} {\bibfnamefont {Z.}~\bibnamefont {Papi{\'c}}},\
  }\href@noop {} {\bibfield  {journal} {\bibinfo  {journal} {Nature Physics}\
  }\textbf {\bibinfo {volume} {17}},\ \bibinfo {pages} {675} (\bibinfo {year}
  {2021})}\BibitemShut {NoStop}%
\bibitem [{\citenamefont {Moudgalya}\ \emph {et~al.}(2021)\citenamefont
  {Moudgalya}, \citenamefont {Bernevig},\ and\ \citenamefont
  {Regnault}}]{moudgalya2021quantum}%
  \BibitemOpen
  \bibfield  {author} {\bibinfo {author} {\bibfnamefont {S.}~\bibnamefont
  {Moudgalya}}, \bibinfo {author} {\bibfnamefont {B.~A.}\ \bibnamefont
  {Bernevig}}, \ and\ \bibinfo {author} {\bibfnamefont {N.}~\bibnamefont
  {Regnault}},\ }\href@noop {} {\bibfield  {journal} {\bibinfo  {journal}
  {arXiv preprint arXiv:2109.00548}\ } (\bibinfo {year} {2021})}\BibitemShut
  {NoStop}%
\bibitem [{\citenamefont {Alhambra}\ \emph {et~al.}(2020)\citenamefont
  {Alhambra}, \citenamefont {Anshu},\ and\ \citenamefont
  {Wilming}}]{PhysRevB.101.205107}%
  \BibitemOpen
  \bibfield  {author} {\bibinfo {author} {\bibfnamefont {A.~M.}\ \bibnamefont
  {Alhambra}}, \bibinfo {author} {\bibfnamefont {A.}~\bibnamefont {Anshu}}, \
  and\ \bibinfo {author} {\bibfnamefont {H.}~\bibnamefont {Wilming}},\ }\href
  {\doibase 10.1103/PhysRevB.101.205107} {\bibfield  {journal} {\bibinfo
  {journal} {Phys. Rev. B}\ }\textbf {\bibinfo {volume} {101}},\ \bibinfo
  {pages} {205107} (\bibinfo {year} {2020})}\BibitemShut {NoStop}%
\bibitem [{\citenamefont {Bernien}\ \emph {et~al.}(2017)\citenamefont
  {Bernien}, \citenamefont {Schwartz}, \citenamefont {Keesling}, \citenamefont
  {Levine}, \citenamefont {Omran}, \citenamefont {Pichler}, \citenamefont
  {Choi}, \citenamefont {Zibrov}, \citenamefont {Endres}, \citenamefont
  {Greiner} \emph {et~al.}}]{bernien2017probing}%
  \BibitemOpen
  \bibfield  {author} {\bibinfo {author} {\bibfnamefont {H.}~\bibnamefont
  {Bernien}}, \bibinfo {author} {\bibfnamefont {S.}~\bibnamefont {Schwartz}},
  \bibinfo {author} {\bibfnamefont {A.}~\bibnamefont {Keesling}}, \bibinfo
  {author} {\bibfnamefont {H.}~\bibnamefont {Levine}}, \bibinfo {author}
  {\bibfnamefont {A.}~\bibnamefont {Omran}}, \bibinfo {author} {\bibfnamefont
  {H.}~\bibnamefont {Pichler}}, \bibinfo {author} {\bibfnamefont
  {S.}~\bibnamefont {Choi}}, \bibinfo {author} {\bibfnamefont {A.~S.}\
  \bibnamefont {Zibrov}}, \bibinfo {author} {\bibfnamefont {M.}~\bibnamefont
  {Endres}}, \bibinfo {author} {\bibfnamefont {M.}~\bibnamefont {Greiner}},
  \emph {et~al.},\ }\href@noop {} {\bibfield  {journal} {\bibinfo  {journal}
  {Nature}\ }\textbf {\bibinfo {volume} {551}},\ \bibinfo {pages} {579}
  (\bibinfo {year} {2017})}\BibitemShut {NoStop}%
\bibitem [{\citenamefont {Bluvstein}\ \emph {et~al.}(2021)\citenamefont
  {Bluvstein}, \citenamefont {Omran}, \citenamefont {Levine}, \citenamefont
  {Keesling}, \citenamefont {Semeghini}, \citenamefont {Ebadi}, \citenamefont
  {Wang}, \citenamefont {Michailidis}, \citenamefont {Maskara}, \citenamefont
  {Ho}, \citenamefont {Choi}, \citenamefont {Serbyn}, \citenamefont {Greiner},
  \citenamefont {Vuletić},\ and\ \citenamefont
  {Lukin}}]{doi:10.1126/science.abg2530}%
  \BibitemOpen
  \bibfield  {author} {\bibinfo {author} {\bibfnamefont {D.}~\bibnamefont
  {Bluvstein}}, \bibinfo {author} {\bibfnamefont {A.}~\bibnamefont {Omran}},
  \bibinfo {author} {\bibfnamefont {H.}~\bibnamefont {Levine}}, \bibinfo
  {author} {\bibfnamefont {A.}~\bibnamefont {Keesling}}, \bibinfo {author}
  {\bibfnamefont {G.}~\bibnamefont {Semeghini}}, \bibinfo {author}
  {\bibfnamefont {S.}~\bibnamefont {Ebadi}}, \bibinfo {author} {\bibfnamefont
  {T.~T.}\ \bibnamefont {Wang}}, \bibinfo {author} {\bibfnamefont {A.~A.}\
  \bibnamefont {Michailidis}}, \bibinfo {author} {\bibfnamefont
  {N.}~\bibnamefont {Maskara}}, \bibinfo {author} {\bibfnamefont {W.~W.}\
  \bibnamefont {Ho}}, \bibinfo {author} {\bibfnamefont {S.}~\bibnamefont
  {Choi}}, \bibinfo {author} {\bibfnamefont {M.}~\bibnamefont {Serbyn}},
  \bibinfo {author} {\bibfnamefont {M.}~\bibnamefont {Greiner}}, \bibinfo
  {author} {\bibfnamefont {V.}~\bibnamefont {Vuletić}}, \ and\ \bibinfo
  {author} {\bibfnamefont {M.~D.}\ \bibnamefont {Lukin}},\ }\href {\doibase
  10.1126/science.abg2530} {\bibfield  {journal} {\bibinfo  {journal}
  {Science}\ }\textbf {\bibinfo {volume} {371}},\ \bibinfo {pages} {1355}
  (\bibinfo {year} {2021})},\ \Eprint
  {http://arxiv.org/abs/https://www.science.org/doi/pdf/10.1126/science.abg2530}
  {https://www.science.org/doi/pdf/10.1126/science.abg2530} \BibitemShut
  {NoStop}%
\bibitem [{\citenamefont {Turner}\ \emph
  {et~al.}(2018{\natexlab{a}})\citenamefont {Turner}, \citenamefont
  {Michailidis}, \citenamefont {Abanin}, \citenamefont {Serbyn},\ and\
  \citenamefont {Papi{\'c}}}]{turner2018weak}%
  \BibitemOpen
  \bibfield  {author} {\bibinfo {author} {\bibfnamefont {C.~J.}\ \bibnamefont
  {Turner}}, \bibinfo {author} {\bibfnamefont {A.~A.}\ \bibnamefont
  {Michailidis}}, \bibinfo {author} {\bibfnamefont {D.~A.}\ \bibnamefont
  {Abanin}}, \bibinfo {author} {\bibfnamefont {M.}~\bibnamefont {Serbyn}}, \
  and\ \bibinfo {author} {\bibfnamefont {Z.}~\bibnamefont {Papi{\'c}}},\
  }\href@noop {} {\bibfield  {journal} {\bibinfo  {journal} {Nature Physics}\
  }\textbf {\bibinfo {volume} {14}},\ \bibinfo {pages} {745} (\bibinfo {year}
  {2018}{\natexlab{a}})}\BibitemShut {NoStop}%
\bibitem [{\citenamefont {Turner}\ \emph
  {et~al.}(2018{\natexlab{b}})\citenamefont {Turner}, \citenamefont
  {Michailidis}, \citenamefont {Abanin}, \citenamefont {Serbyn},\ and\
  \citenamefont {Papi\ifmmode~\acute{c}\else \'{c}\fi{}}}]{PhysRevB.98.155134}%
  \BibitemOpen
  \bibfield  {author} {\bibinfo {author} {\bibfnamefont {C.~J.}\ \bibnamefont
  {Turner}}, \bibinfo {author} {\bibfnamefont {A.~A.}\ \bibnamefont
  {Michailidis}}, \bibinfo {author} {\bibfnamefont {D.~A.}\ \bibnamefont
  {Abanin}}, \bibinfo {author} {\bibfnamefont {M.}~\bibnamefont {Serbyn}}, \
  and\ \bibinfo {author} {\bibfnamefont {Z.}~\bibnamefont
  {Papi\ifmmode~\acute{c}\else \'{c}\fi{}}},\ }\href {\doibase
  10.1103/PhysRevB.98.155134} {\bibfield  {journal} {\bibinfo  {journal} {Phys.
  Rev. B}\ }\textbf {\bibinfo {volume} {98}},\ \bibinfo {pages} {155134}
  (\bibinfo {year} {2018}{\natexlab{b}})}\BibitemShut {NoStop}%
\bibitem [{\citenamefont {Choi}\ \emph {et~al.}(2019)\citenamefont {Choi},
  \citenamefont {Turner}, \citenamefont {Pichler}, \citenamefont {Ho},
  \citenamefont {Michailidis}, \citenamefont {Papi\ifmmode~\acute{c}\else
  \'{c}\fi{}}, \citenamefont {Serbyn}, \citenamefont {Lukin},\ and\
  \citenamefont {Abanin}}]{PhysRevLett.122.220603}%
  \BibitemOpen
  \bibfield  {author} {\bibinfo {author} {\bibfnamefont {S.}~\bibnamefont
  {Choi}}, \bibinfo {author} {\bibfnamefont {C.~J.}\ \bibnamefont {Turner}},
  \bibinfo {author} {\bibfnamefont {H.}~\bibnamefont {Pichler}}, \bibinfo
  {author} {\bibfnamefont {W.~W.}\ \bibnamefont {Ho}}, \bibinfo {author}
  {\bibfnamefont {A.~A.}\ \bibnamefont {Michailidis}}, \bibinfo {author}
  {\bibfnamefont {Z.}~\bibnamefont {Papi\ifmmode~\acute{c}\else \'{c}\fi{}}},
  \bibinfo {author} {\bibfnamefont {M.}~\bibnamefont {Serbyn}}, \bibinfo
  {author} {\bibfnamefont {M.~D.}\ \bibnamefont {Lukin}}, \ and\ \bibinfo
  {author} {\bibfnamefont {D.~A.}\ \bibnamefont {Abanin}},\ }\href {\doibase
  10.1103/PhysRevLett.122.220603} {\bibfield  {journal} {\bibinfo  {journal}
  {Phys. Rev. Lett.}\ }\textbf {\bibinfo {volume} {122}},\ \bibinfo {pages}
  {220603} (\bibinfo {year} {2019})}\BibitemShut {NoStop}%
\bibitem [{\citenamefont {Bull}\ \emph {et~al.}(2020)\citenamefont {Bull},
  \citenamefont {Desaules},\ and\ \citenamefont {Papi\ifmmode~\acute{c}\else
  \'{c}\fi{}}}]{PhysRevB.101.165139}%
  \BibitemOpen
  \bibfield  {author} {\bibinfo {author} {\bibfnamefont {K.}~\bibnamefont
  {Bull}}, \bibinfo {author} {\bibfnamefont {J.-Y.}\ \bibnamefont {Desaules}},
  \ and\ \bibinfo {author} {\bibfnamefont {Z.}~\bibnamefont
  {Papi\ifmmode~\acute{c}\else \'{c}\fi{}}},\ }\href {\doibase
  10.1103/PhysRevB.101.165139} {\bibfield  {journal} {\bibinfo  {journal}
  {Phys. Rev. B}\ }\textbf {\bibinfo {volume} {101}},\ \bibinfo {pages}
  {165139} (\bibinfo {year} {2020})}\BibitemShut {NoStop}%
\bibitem [{\citenamefont {Hudomal}\ \emph {et~al.}(2022)\citenamefont
  {Hudomal}, \citenamefont {Desaules}, \citenamefont {Mukherjee}, \citenamefont
  {Su}, \citenamefont {Halimeh},\ and\ \citenamefont
  {Papi\ifmmode~\acute{c}\else \'{c}\fi{}}}]{PhysRevB.106.104302}%
  \BibitemOpen
  \bibfield  {author} {\bibinfo {author} {\bibfnamefont {A.}~\bibnamefont
  {Hudomal}}, \bibinfo {author} {\bibfnamefont {J.-Y.}\ \bibnamefont
  {Desaules}}, \bibinfo {author} {\bibfnamefont {B.}~\bibnamefont {Mukherjee}},
  \bibinfo {author} {\bibfnamefont {G.-X.}\ \bibnamefont {Su}}, \bibinfo
  {author} {\bibfnamefont {J.~C.}\ \bibnamefont {Halimeh}}, \ and\ \bibinfo
  {author} {\bibfnamefont {Z.}~\bibnamefont {Papi\ifmmode~\acute{c}\else
  \'{c}\fi{}}},\ }\href {\doibase 10.1103/PhysRevB.106.104302} {\bibfield
  {journal} {\bibinfo  {journal} {Phys. Rev. B}\ }\textbf {\bibinfo {volume}
  {106}},\ \bibinfo {pages} {104302} (\bibinfo {year} {2022})}\BibitemShut
  {NoStop}%
\bibitem [{\citenamefont {Else}\ \emph {et~al.}(2016)\citenamefont {Else},
  \citenamefont {Bauer},\ and\ \citenamefont {Nayak}}]{PhysRevLett.117.090402}%
  \BibitemOpen
  \bibfield  {author} {\bibinfo {author} {\bibfnamefont {D.~V.}\ \bibnamefont
  {Else}}, \bibinfo {author} {\bibfnamefont {B.}~\bibnamefont {Bauer}}, \ and\
  \bibinfo {author} {\bibfnamefont {C.}~\bibnamefont {Nayak}},\ }\href
  {\doibase 10.1103/PhysRevLett.117.090402} {\bibfield  {journal} {\bibinfo
  {journal} {Phys. Rev. Lett.}\ }\textbf {\bibinfo {volume} {117}},\ \bibinfo
  {pages} {090402} (\bibinfo {year} {2016})}\BibitemShut {NoStop}%
\bibitem [{\citenamefont {Yao}\ \emph {et~al.}(2017)\citenamefont {Yao},
  \citenamefont {Potter}, \citenamefont {Potirniche},\ and\ \citenamefont
  {Vishwanath}}]{PhysRevLett.118.030401}%
  \BibitemOpen
  \bibfield  {author} {\bibinfo {author} {\bibfnamefont {N.~Y.}\ \bibnamefont
  {Yao}}, \bibinfo {author} {\bibfnamefont {A.~C.}\ \bibnamefont {Potter}},
  \bibinfo {author} {\bibfnamefont {I.-D.}\ \bibnamefont {Potirniche}}, \ and\
  \bibinfo {author} {\bibfnamefont {A.}~\bibnamefont {Vishwanath}},\ }\href
  {\doibase 10.1103/PhysRevLett.118.030401} {\bibfield  {journal} {\bibinfo
  {journal} {Phys. Rev. Lett.}\ }\textbf {\bibinfo {volume} {118}},\ \bibinfo
  {pages} {030401} (\bibinfo {year} {2017})}\BibitemShut {NoStop}%
\bibitem [{\citenamefont {Maskara}\ \emph {et~al.}(2021)\citenamefont
  {Maskara}, \citenamefont {Michailidis}, \citenamefont {Ho}, \citenamefont
  {Bluvstein}, \citenamefont {Choi}, \citenamefont {Lukin},\ and\ \citenamefont
  {Serbyn}}]{PhysRevLett.127.090602}%
  \BibitemOpen
  \bibfield  {author} {\bibinfo {author} {\bibfnamefont {N.}~\bibnamefont
  {Maskara}}, \bibinfo {author} {\bibfnamefont {A.~A.}\ \bibnamefont
  {Michailidis}}, \bibinfo {author} {\bibfnamefont {W.~W.}\ \bibnamefont {Ho}},
  \bibinfo {author} {\bibfnamefont {D.}~\bibnamefont {Bluvstein}}, \bibinfo
  {author} {\bibfnamefont {S.}~\bibnamefont {Choi}}, \bibinfo {author}
  {\bibfnamefont {M.~D.}\ \bibnamefont {Lukin}}, \ and\ \bibinfo {author}
  {\bibfnamefont {M.}~\bibnamefont {Serbyn}},\ }\href {\doibase
  10.1103/PhysRevLett.127.090602} {\bibfield  {journal} {\bibinfo  {journal}
  {Phys. Rev. Lett.}\ }\textbf {\bibinfo {volume} {127}},\ \bibinfo {pages}
  {090602} (\bibinfo {year} {2021})}\BibitemShut {NoStop}%
\bibitem [{\citenamefont {Desaules}\ \emph {et~al.}(2022)\citenamefont
  {Desaules}, \citenamefont {Bull}, \citenamefont {Daniel},\ and\ \citenamefont
  {Papi\ifmmode~\acute{c}\else \'{c}\fi{}}}]{PhysRevB.105.245137}%
  \BibitemOpen
  \bibfield  {author} {\bibinfo {author} {\bibfnamefont {J.-Y.}\ \bibnamefont
  {Desaules}}, \bibinfo {author} {\bibfnamefont {K.}~\bibnamefont {Bull}},
  \bibinfo {author} {\bibfnamefont {A.}~\bibnamefont {Daniel}}, \ and\ \bibinfo
  {author} {\bibfnamefont {Z.}~\bibnamefont {Papi\ifmmode~\acute{c}\else
  \'{c}\fi{}}},\ }\href {\doibase 10.1103/PhysRevB.105.245137} {\bibfield
  {journal} {\bibinfo  {journal} {Phys. Rev. B}\ }\textbf {\bibinfo {volume}
  {105}},\ \bibinfo {pages} {245137} (\bibinfo {year} {2022})}\BibitemShut
  {NoStop}%
\end{thebibliography}%
%%%%%%%%%%%%%%%%%%%%%%%%%%%%%%%%%%%%%%%%%%%%%%%%%%%%%%

%\bibliography{biblio-TI-SSL}
%\bibliography{ref}

%%%%%%%%%%%%%%%%%%%%%%%%%%%%%%%%%%%%%%%%%%%%%%%%%%%%%%
% Appendices
%%%%%%%%%%%%%%%%%%%%%%%%%%%%%%%%%%%%%%%%%%%%%%%%%%%%%%
%\appendix
\end{document}